# A semiconductor-based neutron detection system for planetary exploration

Alejandro Soto[a*], Ryan G. Fronk[b,c], Kerry Neal[a], Bent Ehresmann[a], Steven L. Bellinger[b,c], Michael Shoffner[a], Douglas S. McGregor[c]

[a]*Southwest Research Institute, 1050 Walnut Street, Suit 300, Boulder, CO 80302, USA.*
*asoto@boulder.swri.edu, kneal@boulder.swri.edu, ehresmann@boulder.swri.edu, shoffner@boulder.swri.edu*

[b]*Radiation Detection Technologies, Inc., 4615 S. Dwight Dr., Manhattan, KS 66502, USA,*
*ryan.fronk@gmail.com, bellinger@radectech.com*

[c]*S.M.A.R.T. Laboratory, Department of Mechanical and Nuclear Engineering, Kansas State University, Manhattan, KS 66506, USA, ryan.fronk@gmail.com, bellinger@radectech.com, mcgregor@k-state.edu*

[*]Corresponding Author: Alejandro Soto, +1 (720) 240-0128, asoto@boulder.swri.edu, URL: http://www.alejandrosoto.net

**Abstract**

We explore the use of microstructured semiconductor neutron detectors (MSNDs) to map the ratio between thermal neutrons and higher energy neutrons. The system consists of alternating layers of modular neutron detectors (MNDs), each comprising arrays of twenty-four MSNDs, and high-density polyethylene moderators (HDPE) with gadolinium shielding to filter between thermal neutrons and higher energy neutrons. We experimentally measured the performance of three different configurations and demonstrated that the sensor system prototypes detect and differentiate thermal and epithermal neutrons. We discuss future planetary exploration applications of this compact, semiconductor-based low-energy neutron detection system.

# 1. Introduction

The search for water has long been an important part of the exploration of planets and airless bodies. The distribution of near subsurface water records the origin and evolution of small planetary bodies and provides an *in situ* resource for human exploration (Jones et al. 1990; Cyr, Sears, and Lunine 1998; Sanders and Larson 2013). Neutron spectroscopy has proven successful in mapping water on various planetary bodies, having flown on several missions, including Mars Odyssey (Boynton et al. 2004), Lunar Prospector (Feldman et al. 1999), the Lunar Reconnaissance Orbiter (Mitrofanov et al. 2008), MESSENGER (Goldsten et al. 2007), and Dawn (Prettyman et al. 2003), delivering unprecedented information about the presence of water on Mars, the Moon, Mercury, and asteroids, respectively. Future deep space applications, including future missions with humans, require neutron detection instruments with less mass and power than previous instruments. To meet these demands, we have developed a semiconductor-based low-energy neutron detection system capable of differentiating thermal and epithermal neutrons. The system uses multiple microstructured semiconductor neutron detectors (MSNDs) arranged into a planar-type detector array (McGregor et al. 2015; Ochs et al. 2018). We experimentally measured the performance of three different configurations of the low-energy neutron detection system for spaceflight application, which resulted in a functioning sensor system prototype that is able to detect and differentiate thermal and epithermal neutrons.

Planets and airless bodies are constantly bombarded by galactic cosmic rays (GCRs). When the planetary atmosphere is sufficiently thin or no atmosphere exists, GCRs of sufficient energy penetrate the planetary surface. The collision of incoming cosmic rays with planetary materials in the near surface of the planet produces many neutrons with energy greater than 10 MeV, among other particles. These spallation neutrons collide with subsurface planetary materials before being captured or escaping from the subsurface into space. The escaping neutron flux can be detected by neutron detectors at the surface of the planetary body or from orbit around the planetary body. Neutrons readily interact with hydrogen due to its large scattering cross section, thus moderating any existing flux of neutrons from the fast neutron energy range (>0.5 MeV) to the lower energy epithermal (0.2 eV – 500 keV) and thermal ranges (≤0.2 eV) (Feldman, Reedy, and McKay 1991; Feldman, Boynton, and Drake 1993). The presence of water in the surface can moderate the fluxes of these epithermal and fast neutrons since they lose energy by elastic scattering. Therefore, surface and subsurface water can be mapped by measuring energy-dependent neutron fluxes generated by GCR fluxes (Feldman, Boynton, and Drake 1993; Feldman et al. 1998).

On average, when a neutron collides with a hydrogen nucleus, half the energy of the neutron is transferred during the collision, thereby moderating the epithermal and fast neutron fluxes. Water contains two hydrogen atoms per molecule; therefore, higher water content reduces epithermal and fast fluxes while increasing the thermal flux (Lawrence et al. 2010). The presence of oxygen, which is often found in most planetary regoliths, can shield deeper layers of hydrogen from detection via the fast neutron flux. Therefore, measuring the epithermal and fast neutron fluxes separately can also provide information about the burial depth of water. All of these energy loss processes are dependent on the depth distribution and content of water in the ground, so that the concurrent measurement of all three neutron energy ranges provides a unique detection method for the presence, depth, and abundance of water (Lawrence et al. 2010).

We are interested in a low-mass, compact neutron detection system that can map the presence of subsurface water on an airless body, either from orbit or *in situ*. Such an instrument can play a significant role in the exploration of airless planetary bodies like the Moon and asteroids (e.g., Feldman et al. 1993; Lawrence et al. 2010; Prettyman et al. 2017). To achieve a low mass and compact configuration, we developed a detection system that strictly maps the existence of subsurface water using the ratio of epithermal (and higher energy) neutrons to thermal neutrons. This method has been used by previous missions, such as the Lunar Prospector mission (Feldman et al. 1998). Using newer, high-efficiency, semiconductor-based neutron detectors, we have developed a more compact neutron detection system for future mapping of subsurface water on planetary bodies.

## 2. Neutron detection methods

A typical method for detecting neutrons uses *gas-filled proportional counters*, which are usually a tube filled with $^3He$ gas (Crane and Baker 1991). Neutrons reacting with the fill gas yield the $^3He(n,p)^3H$ reaction, which in turn ionizes the gas. Charges are drifted to the detector electrodes and read-out as voltage pulses, and recorded in a pulse-height spectrum to yield neutron counts. $^3He$ counters have a very high efficiency to detect thermal neutrons (commonly 70% efficiency) and show a good discrimination of gamma-ray signals. The major drawback for this technology is the price and rarity of $^3He$. Furthermore, commonly used $^3He$ proportional counters require high-voltage operations and are comparably large in volume and mass. Alternatively, $BF_3$-filled gas tubes have been used. $BF_3$ is more economical than $^3He$, but has a lower detection efficiency (about half as efficient). Also, $BF_3$ is highly toxic, making it difficult and expensive to use and generally $BF_3$ counters have the same mass and bulk of $^3He$ counters.

Some neutron detection systems use *plastic or liquid scintillators*. A scintillator is a material that fluoresces when struck by an energetic charged particle, neutron, or gamma ray. Neutrons interact with the nuclei of the scintillator material by elastic scattering, creating 'recoil protons'. These recoil protons further interact with the scintillator material and produce light pulses, which can be read out by a photo-detector. The dependence of the amplitude of the recoil pulse on the incident neutron energy is well known, enabling the determination of the energy of sufficiently energetic neutrons. However, spectroscopy is complicated by multiple nonlinear effects, especially in the presence of a broad spectrum of neutron energies. They have very good detection efficiency, but high gamma-ray sensitivity, so that the differentiation between neutrons and background cannot necessarily be determined by the pulse height information alone, unless the scintillator also contains $^{10}B$ or $^6Li$ and the capture-gated method is employed, or a method of pulse shape discrimination is used, either of which adds considerable complexity to the instrument. Organic scintillators are insensitive to thermal neutrons, mainly because the scintillation mechanism relies on excitation by energetic recoil protons.

Finally, *semiconductor detectors* (for example Si-based) coated with a thin-film of neutron reactive material (e.g., $^{10}B$ or $^6Li$) have been studied for a long time. A neutron that interacts with the coating material creates energetic reaction products that can enter the semiconductor diodes and be read-out as pulse-height spectra. The detection efficiency for thermal neutrons is typically low compared to gas-filled proportional counters. However, the major advantages of such a semiconductor neutron detector are its low power consumption, its compactness, and low cost to produce.

Due to the advantages outlined in the section above, the use of semiconductor-based neutron detectors for planetary neutron spectroscopy is desirable. Previous designs for semiconductor neutron detectors have low efficiency (McGregor et al. 2003). However, recent progress has been achieved using microstructured semiconductor neutron detectors (MSNDs) (McGregor, Bellinger, and Shultis 2013). MSNDs have small channels etched into the surface semiconductor substrate, which are backfilled with a neutron-reactive material. Using this method, the thermal-neutron detection efficiency for single-sided MSNDs can be raised to >30%, and over 65% for double-sided MSNDs, thereby making this a potential detector technology for planetary neutron spectroscopy (Bellinger et al. 2013; Ochs et al. 2018). The detector consists of a Si diode with etched channels backfilled with nano-sized $^6LiF$ powder. The absorption of a neutron by $^6LiF$ creates a ~2.73 MeV triton and a ~2.05 MeV alpha particle that can be detected with the Si diode. Shown in Figure 1 is a simulated neutron response of an MSND over the thermal, epithermal, and fast neutron energy ranges. For the low-energy neutron detection system, we used modular neutron detector (MND) boards, developed by Radiation Detection Technologies, Inc. (RDT), each of which holds 24 MSNDs (Ochs et al. 2019). The MND layout increases the cross-sectional area available for neutron detection. Figure 2 shows an MND board with the 4x6 array of MSNDs.

We designed and built a prototype low-energy neutron detection system consisting of multiple MNDs for planetary remote sensing applications. Our design is similar in technique to portable neutron spectrometers for human dose equivalence estimation proposed by Oakes et al. (2013) and Hoshor et al. (2015). Our design differs from the Oakes et al. (2013) and Hoshor et al. (2015) designs by drastically reducing the number of MSND layers and adjusting the moderator thicknesses in order to maximize the detections used in the ratio calculations.

The MSND detectors detect neutrons across a broad range of energies (see Figure 1). To use these detectors to map subsurface hydrogen and water in planetary bodies, such as asteroids, we have developed an instrument design that separates the thermal neutron signal from higher neutron energy signals. The MSND-based instrument can then map hydrogen and water by measuring the ratio of thermal neutrons to higher energy neutrons. This neutron detection technique was used in early space-based neutron detectors, albeit with much larger detection devices, including scintillators (Feldman et al., 1991; Feldman et al., 2000). We explored three configurations of this instrument, where we tried to maximize the detection of thermal and higher energy neutrons. The resulting design is a low-energy neutron detection system that demonstrates the capability of differentiating hydrogen abundance in an asteroid or other planetary environment.

## 3. Instrument Design

We developed and tested a number of detector assembly configurations of the low-energy neutron detection system. At a minimum, we wanted a detector configuration that differentiates the thermal neutrons from the higher energy neutrons. Such a configuration can be used to detect the presence of subsurface water on a planetary surface. However, we also tested two more configurations, in which we attempted to increase the information about higher energy neutrons that the instrument can acquire. Table 1 provides details about the various configurations while Figures 3, 4, and 5 show the primary components of the configurations.

The simplest configuration, Configuration #1, uses two MNDs separated by a thin sheet of gadolinium (Gd), as shown schematically in Figure 3. Since $^{157}$Gd has the largest known cross section for thermal neutrons with an area of 2.5 X $10^5$ barns[1] (Leinweber et al. 2006), then each 1.27 millimeter thick sheet of $^{157}$Gd captures >99% of the incident thermal neutrons, making such a foil an ideal thermal neutron shield. With this configuration, the combined flux of epithermal and fast neutrons can be separated from the flux of thermal neutrons. The Gd shields in Configurations #2 and #3 have the same thickness and provide the same function.

With Configuration #1 (see Figure 3), the MND 1 detects all of the incoming neutrons, weighted by the detector responsivity (Figure 1). The MND 2 detects only those neutrons with energies above the Gd cutoff, which is ~0.5 eV, since the Gd absorbs nearly 100% of the neutrons below the cutoff (D'Mellow et al. 2007). Thus, the simplest ratio that we can measure is the neutron flux at MND1 versus the neutron flux at MND2, i.e.,

$$\frac{F_{ef}}{F_{total}} = \frac{F_{MND2}}{F_{MND1}}, \quad (1)$$

where $F_{ef}$ is the flux of epithermal and fast neutrons that pass through the *Gd* shield. The thermal neutron flux, $F_{th}$, is then

$$F_{th} = F_{total} - F_{ef}. \quad (2)$$

Thus, we can derive an epithermal to thermal neutron flux ratio $F_{ef}/F_{th}$, where we assume that the bulk of the $F_{ef}$ flux is due to epithermal neutrons due to the responsivity function of the MSND detectors), which has been shown to be diagnostic of the hydrogen and water content of airless planetary bodies (Feldman et al. 1991; Lawrence et al. 2010).

We experimentally evaluated two more configurations. In both of these additional configurations we used high-density polyethylene (HDPE) to moderate the epithermal and fast neutron energies (Hoshor et al. 2015). If we place the HDPE after an MND, then some of the neutrons that were not detected will be backscattered and may be detected in a second pass through the MND. Thus, the backscatter properties of the HDPE increases the intrinsic detection efficiency (Hoshor et al. 2015; Kittelmann et al. 2017). Additionally, the HDPE moderates the neutrons, moving the energy of the scattered neutron towards the peak detection efficiency of the MSNDs, which occurs at lower energies (see Figure 1). Therefore, if we place a sheet of HDPE after the Gd shield, then any epithermal neutrons that make it past the Gd shield are more likely to be detected because their energies will be moderated to lower energy levels (i.e., into the thermal energy regime) where the MND detectors are more likely to detect incident neutrons. This moderation process increases the detector response at higher initial energies thereby amplifying the detection of weak neutron signals.

---

[1] A barn (symbol b) is a unit of area equal to $10^{-24}$ cm$^2$.

*Table 1. Test configurations for the thermal and epithermal neutron tests at KSU. The configuration column uses the following codes: D = MND board; G = gadolinium; P = HDPE (i.e., polyethylene); B = boron-loaded HDPE.*

| Configuration # | Configuration | PE thicknesses (in) |
|---|---|---|
| 1 | D-G-D | N/A |
| 2 | D-G-P-D-P-G-B | 0.75, 0.75, 2 |
| 3 | D-G-P-D-P-G-P-D-P | 0.75, 0.75, 1.0, 1.0 |

Configuration #2 is shown in Figure 4. In this configuration, we used HDPE to both backscatter and moderate the higher energy neutrons. Two types of HDPE are used: standard, undoped HDPE and boron-doped HDPE. MND 1 measures the incoming neutron flux (primarily in the thermal neutron energy range). The thermal neutrons are absorbed by the first *Gd* shield, while the higher energy neutrons largely pass through the first *Gd* shield. These higher energy neutrons then encounter the first HDPE block. Some of the neutrons that encounter the HDPE block are scattered back towards MND 1, where a fraction of these neutrons are detected by MND 1. Some of the other higher energy neutrons that remain are moderated to thermal energies as they pass through the HDPE and some subset of the original neutrons screened by the *Gd* pass through the HDPE unchanged. MND 2 will detect this resulting neutron flux, whose energy distribution is a mix of the original higher-energy neutrons and the newly moderated thermal neutrons. Any neutrons that pass through MND 2 without detection will encounter another block of HDPE, which may backscatter and moderate some of the neutrons. Again, any thermalized neutrons that are not backscattered are then filtered by the second Gd shield.

At this point, the neutrons encounter borated-HDPE, which in our experiments substituted for a boron-loaded scintillator for detecting fast neutrons (for example, the FND used in the ISS-RAD instrument; see Leitgab et al, 2016 and Zeitlin, 2013). There are various applications where the addition of a fast neutron detector is beneficial (see the Discussion section below), therefore, Configuration #2 explored the impact of such a detector on our instrument system.

Calculating the $F_{ef}/F_{th}$ flux ratio for Configuration #2 involves the same process as in Configuration #1. The advantage of Configuration #2 is the detection of additional higher energy neutrons that have been moderated to energies where the detectors have greater responsivity. Configuration #2 depends more on post-processing modeling for interpretation of the results, but it improves the signal detection of the instrument.

Configuration #3, shown in Figure 5, is similar configuration #2, except the borated HDPE is replaced with another set of HDPE blocks and an additional MSND detector board (i.e., $B$ is replaced with a P-D-P sequence). In this configuration, to calculate the $F_{ef}/F_{th}$ flux ratio, the neutron detections at MND 2 and MND 3 are combined to give a total $F_{ef}$ flux. This configuration is an attempt to maximize the number of neutrons detected from the original flux by using two MNDs after the $Gd$ shield to maximize the detection of the higher energy neutrons. Similar to Configuration #2, Configuration #3 depends on post-processing modeling for interpretation of the results improving the detection of the higher-energy neutrons.

## 4. Fabrication

A prototype instrument was developed that allowed us to quickly change the configuration during testing. As shown in Figure 6, the neutron detection system includes a chassis in which the MNDs, the HDPE sheets, and the $Gd$ plates are stacked in a variety of configurations. The top and back remained open during the operation of the low-energy neutron detection system. The chassis is made of aluminum, minimizing interactions with the neutrons and gamma rays generated during experimentation.

For instrument readout, the electronics interface is based on previous work with the MND technology (Ochs et al 2019). When a neutron is absorbed in the neutron-converting material, the resulting charged-particle reaction products induce excitation within the MSND diode volume, and electron-hole pairs are produced. These electron-holes pairs drift to their respective electrodes, thereby producing a current that charges a capacitor. The potential 'pulse' developed on the capacitor is measured and amplified by the readout electronics of the MND. The amplitude of the pulse produced is directly proportional to the original amount of energy deposited into the active diode region of the semiconductor volume. Pulses are therefore either 'accepted' or 'rejected' based on their amplitude, which is often indicative of the origin of the energy deposition; pulses generated by thermal noise or gamma-ray interactions within the diode are generally much smaller in magnitude that those from neutron-capture-induced charged-particle reaction product interactions. A lower-level discriminator (LLD) value is established to reject pulses generated from gamma-ray and low-level thermal noise current.

Pulses exceeding the LLD setting are 'accepted', at which point the discriminator electronics circuit produces a digital square-wave pulse with a magnitude of 5 volts, with a pulse width of between 5-50 μs, depending on the time spent above the threshold value. The digital pulse is passed to an output driver board which drives the pulse at 5-volts and 50-Ω impedance via a subminiature version 'A' connector (SMA) connector. The signal can then be read by any digital-sensing electronics, such as a counter-timer or digital acquisition board.

## 5. Development of Models and Model Validation

Accurate theoretical models of the detector system must be developed in order to validate measured results and to predict how possible changes to the design may affect system performance. The present neutron-energy spectrometer was modeled using a combination of custom Python scripts and the Monte Carlo N-Particle (MCNP6) code (Goorley et al. 2012). Accurate and detailed models of the problem geometry can be created and analyzed. Generally speaking, MCNP6 uses Monte Carlo methods to accurately reproduce the emissions of a defined radiation source, transport the source particles through the problem geometry (while modeling all methods of interaction with the surrounding media), and, if necessary, model a radiation detector's response to the capture and measurement of said particle (Goorley et al. 2012). In the present work, the entire neutron-energy spectrometer assembly was modeled, including all sensors and materials used in the fabrication of the instrument.

### 5.1. Environmental Modeling

The geometry used in the simulations presented here was modeled after the real-world location used for all detector-source measurements. The room was modeled as 7.3-m long, by 4.5-m wide, by 4.25-m high (inner dimensions), with a 0.25-m thick concrete wall in all directions (Figure 7). The density of the concrete was assumed to be 2.31 g cm$^{-2}$. The air within the virtual room was filled with air at 1-atm at 0% relative humidity, though the humidity and atmospheric pressure was likely different. In order to simplify the model, no equipment or personnel present during the real-world measurements were modeled for this problem. The detector assembly and radiation sources were modeled approximately where they were located during the real-world measurements. The detector assembly was mounted onto borated HDPE (see Figure 6, bottom, and Figure 7, right) and placed on a cart approximately 107 cm (z-direction) from the floor (see Figure 6). The detector assembly was centered in the room in the y-direction and located 280 cm from the nearest wall in the x-direction. Neutrons scattering out of the concrete walls and leaving the room were terminated and no longer considered in the problem. Borated HDPE was placed around the experimental detector apparatus in an attempt to minimize neutron albedo contributions from the nearby walls and floors, which was recreated in the model. Because the borated HDPE does affect neutron transport considerably from the source to the detector through neutron scattering, it was therefore included in the model.

### 5.2. Neutron Sensor Modeling

The core neutron sensor technology, the microstructured semiconductor neutron detector (MSND), was modeled as $^6LiF$ rectangular parallel piped trenches embedded in an $Si$ diode. The trenches were backfilled with 95%-enriched $^6LiF$, and the trenches were 20-μm wide and 300-μm deep, with an overall density of 0.892 g cm$^{-2}$. Crystalline density of $^6LiF$ is 2.55 g cm$^{-3}$, however, the method by which the $^6LiF$ powder is filled into the microcavities, a packing fraction of approximately 35% is achieved. The backfilled trenches are repeated every 30 μm, with the remaining volume filled with *Si*, as shown in Figure 8 (Shultis and McGregor, 2009; McGregor et al., 2015). Neutrons intersecting the $^6LiF$ trenches have some probability of absorption by the $^6Li$ nucleus based on the angle of impact and the neutron absorption cross section for the given energy of the neutron. Upon absorption, the system undergoes the $^6Li(n,t)^4He$ reaction, in which the charged-particle reaction products are then transported in opposite directions from each other at some randomly-selected trajectory. Particle energy deposition is tracked while transporting the charged particles, and any energy deposited into the *Si* fins is tallied using an F8 tally. The event produces a 'count' if the energy deposition exceeds the LLD value, calibrated at 20% intrinsic thermal neutron detection efficiency as was measured in the real-world calibration procedure (McGregor and Shultis 2011.).

## 5.3. Modular Neutron Detector Modeling

Each Modular Neutron Detector (MND) consists of twenty-four MSNDs arranged in 4 x 6 array (Figure 9). Each MSND is contained within a ceramic ($Al_2O_3$) detector board (CDB) and attached to a common FR4 electronics board. The CDB is then surrounded with a mu-metal electromagnetic (EM) shield. The backside of the FR4 electronics board is layered with additional EM shielding. The back-side supporting electronics were omitted.

## 5.4. Detector Assembly

The neutron spectrometer assembly comprises four primary components, in quantities depending on the detector scenario: MNDs, HDPE neutron moderator, Gd thermal neutron attenuators, and the aluminum supporting structure. Much of the aluminum supporting material was omitted from the MCNP models due to the low probability of interaction of neutrons over a wide range of energies. However, the front, side, and bottom aluminum plates were considered in the models.

There were three primary detector configurations considered, as shown in Table 1. Each configuration was mounted in an instrument assembly built of aluminum with a borated HDPE backplate. The model depiction of Configuration #1, which is a 2-Detector assembly consisted of an MND-Gd (thermal-neutron shielding layer)-MND, is shown in Figure 10, including the additional instrument material, primarily aluminum that was accounted for in the modeling. Similarly, shown in Figure 11 is a model depiction of Configuration #2, where the HDPE between the MNDs are 1.905 cm thick. A block of borated HPDE is used as a stand-in for a fast neutron detector. Finally, Figure 12 shows how Configuration #3 was depicted in the model, where the first set of HDPE layers had a thickness of 1.905 cm while the last set measured 2.54 cm thick.

## 5.5. Radiation Source Modeling

Two separate radiation sources were modeled to compare to the real-world measurements performed: a $^{252}Cf$ source and an AmBe source. The $^{252}Cf$ source was approximated as a point isotropic source located centrally inside of a 304L stainless steel source enclosure.

The outer dimensions of the enclosure measured 0.94 cm in diameter and 3.475 cm tall. The acrylic carrying case surrounding the stainless steel was also modeled and had a thickness of 2.5 mm. Neutrons from the $^{252}Cf$ source were emitted uniformly in all directions and were assigned energies based on the Watt fission spectrum using constants a = 1.18 and b = 1.03419 (Figure 13).

Determination of the *AmBe* source energy distribution was less straightforward than with the $^{252}Cf$ source. The energy distribution from the ($\alpha$, n) reaction within a given *AmBe* source is not well defined and can vary from experimental source to source depending on the *Am* compound used, the method of mating the *Am* to the *Be*, density, etc. Therefore, the source energy distribution was determined using SOURCES4C which models the ($\alpha$, n) reaction assuming that the *Am* compound and *Be* are homogenously distributed in a given volume. The source used for the real-world measurements was composed of $AmO_2$ sintered to Be to form $AmO_2Be_{19}$. The output spectrum predicted by SOURCES4C is found in Figure 12. The active AmBe element was assumed to be a point isotropic source encased in 304L stainless steel, measuring 1.90 cm tall with a diameter of 1.415 cm.

Both the $^{252}Cf$ and *AmBe* sources were also measured while in a HDPE moderator cylinder cask for the 'moderated' tests. These cylinders were modeled as 1.00 g cm$^{-3}$ HDPE, 5.45 cm tall and approximately 9.7 cm diameter. An air cavity was modeled in the center of the casks roughly the dimensions of the aforementioned sources. The sources were located at the bottom of the cavity, and the cavity was placed at the appropriate distance from the detector assemblies.

# 6. Results

## 6.1. Experiment setup

We conducted tests of the low-energy neutron detection system at Kansas State University using two well-calibrated and standard neutron sources (acquired from Frontier Technology Corporation), to provide thermal and epithermal neutrons to our instrument. We used a radioactive californium ($^{252}Cf$) source, which has a peak neutron energy of 9.0 x 10$^5$ eV. Fortunately, the $^{252}Cf$ energy spectrum is similar to a standard fission spectrum, covering the lower range of epithermal neutrons. We used americium mixed with beryllium (*AmBe*), as described in the previous section, which allowed us to probe the higher end of the epithermal energy range. For all of the tests, we used the three different configurations of the MSNDs within the low-energy neutron detection system, as discussed above and shown in Table 1. Additionally, we tested the gamma-ray rejection performance of the low-energy neutron detection system. We continually took background measurements to assess the neutron contribution from the testing environment. The exposure times were selected to achieve standard deviations that are on the order of 1% to 2% of the total counts for each experiment.

Some of the experiments included sources that were "moderated". In these experiments, the $^{252}Cf$ and *AmBe* sources were surrounded by a cylinder of HDPE in order to increase the number of lower-energy neutrons impinging on the detector setup, as described in the previous section. The drawback to this "moderated" source is that the neutron energies are spread over a larger range. However, the wider range of energies provided by the moderated sources better simulate natural sources of neutrons.

## 6.2. Experiment Results

As discussed above, the first MND in any configuration detects primarily thermal and epithermal neutrons due to the detector response function (Figure 1). In all of the configurations, the second MND responds to any of the higher energy neutrons that were not shielded by the Gd plate. In configurations #2 and #3, the second MND measures epithermal and fast neutrons, some of which are moderated to thermal energies. Finally, for configuration #3, the third MND measures some of the higher energy neutrons that were not shielded, detected, nor moderated by previous components in this configuration. Listed in Table 2 are the data collected with our experiments, in units of neutron counts per second. The twelve experiments cover all three configurations with both sources. Each source is exposed to the instrument in both a bare and a moderated configuration (discussed above). The standard deviation for the count rates, shown in Table 2, was calculated as

$$\sigma = \sqrt{(n_s + n_b)/T_t + n_b/T_b} \,, \tag{3}$$

where $n_s$ is the number of neutrons counted when exposed to the source, $n_b$ is the number of background neutrons counted, $T_b$ is the background exposure time, and $T_t$ is the total exposure time ( experiment and background exposure time combined).

*Table 2. List of conducted experiments describing instrument and neutron source configurations, as well as detector counts per second (cps).*

| Experiment | Config. | Source | Source Condition | Detector #1 (cps) | Detector #2 (cps) | Detector #3 (cps) |
|---|---|---|---|---|---|---|
| 1 | 1 | Cf-252 | Bare | 1.26 ± 0.01 | 0.82 ± 0.01 | - |
| 2 | 1 | Cf-252 | Moderated | 30.25 ± 0.03 | 2.83 ± 0.01 | - |
| 3 | 1 | AmBe | Bare | 3.19 ± 0.01 | 2.25 ± 0.01 | - |
| 4 | 1 | AmBe | Moderated | 40.13 ± 0.03 | 4.87 ± 0.01 | - |
| 5 | 2 | Cf-252 | Bare | 2.16 ± 0.02 | 16.49 ± 0.05 | - |
| 6 | 2 | Cf-252 | Moderated | 25.25 ± 0.06 | 18.75 ± 0.05 | - |
| 7 | 2 | AmBe | Bare | 3.61 ± 0.02 | 21.77 ± 0.05 | - |
| 8 | 2 | AmBe | Moderated | 42.67 ± 0.06 | 38.16 ± 0.06 | - |
| 9 | 3 | Cf-252 | Bare | 2.07 ± 0.002 | 16.45 ± 0.004 | 7.62 ± 0.003 |
| 10 | 3 | Cf-252 | Moderated | 23.06 ± 0.003 | 17.70 ± 0.003 | 4.45 ± 0.001 |
| 11 | 3 | AmBe | Bare | 3.46 ± 0.02 | 21.42 ± 0.05 | 11.54 ± 0.04 |
| 12 | 3 | AmBe | Moderated | 42.50 ± 0.06 | 38.44 ± 0.06 | 11.54 ± 0.06 |

Shown in Figure 14 are the results from our experiments using configuration #1. The plot shows the count rate measured by each detector for each of the sources used. When measuring neutrons from the unmoderated $^{252}Cf$ source, both the MND 1 and the MND 2 measure low rates of neutrons. The signal from the moderated source, however, is an order of magnitude larger for the first detector, which is due to the spectral shift created by the HDPE cylinder. The neutron count rate at the second detector, however, does not increase significantly for the moderated source. Basically, very few neutrons with sufficiently low energy to be detectable by the MND are transmitted through the *Gd* shield.

Configuration #2 results are shown in Figure 15. The neutron detection rates in the first detector (MND 1) in Figure 15 are similar to the results from Figure 14 because the subsequent sequence of detectors and HDPE blocks in Configuration #2 have no effect on the flux of neutrons at the first detector. However, the MND 2 detector reveals a different neutron response compared to the response in the same detector in configuration #1. The MND 2 detected an augmented signal due to the HDPE block placed after the *Gd* shield in the instrument, in both the unmoderated and moderated source test. Any neutrons that streamed through the *Gd* shield interacted with the HDPE, leading to moderation of the high energies of these neutrons and thus increasing the detection rate in the MND 2.

In Figure 16, we directly compare the results from Configuration #1 and #2. From Figure 16 it is shown that the two configurations measure essentially the same neutron count rate in the first detector, MND 1. The additional HDPE material used in Configuration #2, however, clearly increases the number of neutrons detected by MN2 for all sources. Thus, Configuration #2 effectively increases the responsivity of the MND 2 by moderating the higher energy neutrons to lower, detectable energies.

Finally, shown in Figure 17 are the count rates for tests using configuration #3. Again, the MND 1 count rates for an unmoderated source are similar to the other configurations. The use of HDPE blocks within the instrument again increases the signal on the MND 2 detector while the moderated source is increasing the signal at both MND 1 and MND 2. The unmoderated signal at a third detector, MND 3, is lower than detected at MND 2 indicating the number of neutrons available to detect has dropped significantly, even after being moderated or backscattered by one of the HDPE blocks surrounding the MND 3. Additionally, a moderated source leads to an even lower detection at MND 3, due to the more thermalized spectrum emitted from the source geometry.

## 6.3.     Comparison of Experiment Data with Model Predictions

The twelve experiments were modeled as described in the previous sections. Tallies for each detector were multiplied by the relative neutron emission rates as assayed by their manufacturer, roughly 58,600 n s$^{-1}$ for the $^{252}Cf$ source, and 192,600 n s$^{-1}$ for the *AmBe* source. Simulation results are listed in Table 3.

*Table 3. Expected neutron count-rates (cps) for the different experimental setups as modeled in MCNP6.*

| Experiment | Config. | Source | Source Condition | Detector #1 (cps) | Detector #2 (cps) | Detector #3 (cps) |
|---|---|---|---|---|---|---|
| 1 | 1 | Cf-252 | Bare | 2.06 ± 0.11 | 1.10 ± 0.08 | - |
| 2 | 1 | Cf-252 | Moderated | 26.12 ± 0.39 | 2.43 ± 0.12 | - |
| 3 | 1 | AmBe | Bare | 3.94 ± 0.28 | 2.42 ± 0.22 | - |
| 4 | 1 | AmBe | Moderated | 42.30 ± 0.91 | 5.03 ± 0.32 | - |
| 5 | 2 | Cf-252 | Bare | 3.06 ± 0.13 | 22.91 ± 0.36 | - |
| 6 | 2 | Cf-252 | Moderated | 27.06 ± 0.40 | 17.87 ± 0.32 | - |
| 7 | 2 | AmBe | Bare | 5.62 ± 0.33 | 38.42 ± 0.87 | - |
| 8 | 2 | AmBe | Moderated | 44.61 ± 0.94 | 46.02 ± 0.95 | - |
| 9 | 3 | Cf-252 | Bare | 3.04 ± 0.13 | 23.04 ± 0.37 | 8.92 ± 0.23 |
| 10 | 3 | Cf-252 | Moderated | 27.08 ± 0.40 | 17.86 ± 0.32 | 3.65 ± 0.15 |
| 11 | 3 | AmBe | Bare | 5.54 ± 0.33 | 39.11 ± 0.88 | 21.15 ± 0.65 |
| 12 | 3 | AmBe | Moderated | 44.63 ± 0.94 | 45.96 ± 0.96 | 13.19 ± 0.51 |

Generally, there was good agreement between the model and the real-world measurements, as shown in Figure 18. The model overestimated the expected count rates for the bare sources, which will require further investigation. The primary goal of the instrument, however, is to determine the ratio of the thermal neutron flux to higher energy neutron fluxes (including epithermal and fast neutron fluxes), and these ratios match well with the real-world data. Shown in Figure 19 are both the predicted and measured ratio of the MND2 detections to the MND 1 detections, for experiments 1-8, and the predicted and measured ratio of the detections measured by MND 2 and MND 3, combined, to the MND 1 detections, for experiments 9-12. This ratio, (MND 2)/(MND 1) or (MN3 2 + MND 3)/(MND 1), is essentially a ratio of the epithermal and fast neutron flux to the thermal neutron flux (i.e., $F_{ef}/F_{th}$), since the detector responsivity to fast neutrons is so low. Here we see a close agreement between the predicted ratio and the measured ratio. For all but two experiments, the predicted $F_{ef}/F_{th}$ ratio and the measured $F_{ef}/F_{th}$ ratio match to within their respective uncertainties (Figure 19). The largest discrepancies occur for the experiments that used a bare *AmBe* source. Regardless, the difference between the predicted $F_{ef}/F_{th}$ ratio and the measured $F_{ef}/F_{th}$ ratio is small enough that the models can be used to predict future performance of this type of instrument.

## 7. Discussion and Conclusion

The modeling and experimental measurements shown above demonstrate that we can measure the ratio of neutrons of different energies, and that we can model the results to a sufficient accuracy to derive the original neutron source, which can be used to accurately surmise the water content of extraterrestrial soil. This process was successful in all three configurations. What differentiates the three configurations is the extent to which the individual configuration maximizes the detection of neutrons with different energies.

Configuration #1 provides the most compact method to simply ascertain the ratio of thermal to higher energy (mainly epithermal) neutron fluxes, which is shown in Experiments 1-4 in Figure 19. For this configuration, the predicted and measured ratios are very close, demonstrating our ability to model the detection process and extract the original neutron signal. Note that the value of the ratios in Experiment 2 is extremely low because the moderated $^{252}Cf$ source created a large number of thermal neutrons, which moved the $F_{ef}/F_{th}$ ratio towards zero.

Configurations #2 and #3 start to show larger mismatches between the predicted and the measured ratios. The overall measured neutron signal, however, is much larger, which might be an advantage depending on the environment being mapped, as necessary integration times for the measured signals to achieve desired statistical certainty will be reduced. Furthermore, further work and calibration of this type of instrument will allow us to better model the response of the instrument, thereby, improving the modeling of the neutron detection process. Regardless of configuration, interpreting the neutron ratios will require modeling of the instrument system, which is typical for space-based instruments that map neutron emission (see, for example, Lawrence et al. 2006).

Because of the difficulty of delivering spacecraft to deep space, such missions are resource-limited, where much of the resources are allocated to the launch system and spacecraft. To maximize the scientific return of such a mission, instruments must deliver the maximum amount of scientific data with the lowest possible requirement for resources, such as mass, volume, and power. For a neutron detection system whose scientific return is defined by its capability to detect the presence of water-equivalent hydrogen, the key metric is the capability to measure the highest possible neutron count rates, $R_N$ (e.g., due to a large detection volume, or a high neutron-detection efficiency), while limiting the demand on spacecraft resources. In designing this neutron detection system, we have striven to minimize the mass and volume of the instrument to maximize the merit of the instrument for a given set of observations.

There are a number of future human and robotics planetary missions that would benefit from this type of instrument. For small satellites, e.g. cubesats, in orbit over airless bodies, like asteroids, our compact neutron detection system can provide a simple method for mapping subsurface moderators, like hydrogen and water. The small mass and volume of our instrument, compared to other neutron detection instruments, allows for the accommodation of the instrument on a small satellite without precluding the use of other instruments. For human exploration of the moon and asteroids, our compact neutron detection system can be used during field surveying, in order to map the possible existence of subsurface water, which would then help drive the planning of ongoing in situ science operations on a human mission to a planetary body.

Our tests demonstrate that the MSND detectors can be used to detect thermal and epithermal neutrons and to differentiate the thermal neutrons from the higher-energy neutrons. With the proper configuration of detectors and moderating/back-scattering material, our system will be capable of mapping subsurface water and other neutron moderators on airless planetary bodies.

## 8. Acknowledgements

Operation of the TRIGA Mk II research reactor located at Kansas State University was performed by Dr. Jeffrey A. Geuther and his reactor operations team and was greatly appreciated. Detectors were designed, fabricated, and characterized at the Semiconductor Materials and Radiological Technologies (S.M.A.R.T.) Laboratory at Kansas State University and at Radiation Detection Technologies, Inc.

Funding: This work was supported by a Southwest Research Institute internal research grant.

Goldsten, John O., Edgar A. Rhodes, William V. Boynton, William C. Feldman, David J. Lawrence, Jacob I. Trombka, David M. Smith, et al. (2007). "The Messenger Gamma-Ray and Neutron Spectrometer." *Space Science Reviews* 131 (1): 339–91.

Goorley, T, M James, T Booth, F Brown, J Bull, LJ Cox, J Durkee, et al. (2012). "Initial Mcnp6 Release Overview." *Nuclear Technology* 180 (3). Taylor & Francis: 298–315.

Hoshor, C.B., T.M. Oakes, E.R. Myers, B.J. Rogers, J.E. Currie, S.M. Young, J.A. Crow, et al. (2015). "A Portable and Wide Energy Range Semiconductor-Based Neutron Spectrometer." *Nuclear Instruments and Methods in Physics Research Section A: Accelerators, Spectrometers, Detectors and Associated Equipment* 803 (December). Elsevier BV: 68–81.

Jones, Thomas D., Larry A. Lebofsky, John S. Lewis, and Mark S. Marley. (1990). "The Composition and Origin of the c, P, and d Asteroids: Water as a Tracer of Thermal Evolution in the Outer Belt." *Icarus* 88 (1): 172–92.

Kittelmann, T., E. Klinkby, X. Xiao Cai, K. Kanaki, C. P Cooper-Jensen, and R. Hall-Wilton. (2017). "Using Back-Scattering to Enhance Efficiency in Neutron Detectors." IEEE Transactions on Nuclear Science, 64, 1562-1573.

Lawrence, D. J., Feldman, W. C., Elphic, R. C., Hagerty, J. J., Maurice, S., McKin- ney, G. W., and Prettyman, T. H. (2006). Improved modeling of Lunar Prospector neutron spectrometer data: Implications for hydrogen deposits at the lunar poles. Journal of Geophysical Research: Planets, 111(E8):n/a–n/a. E08001.

Lawrence, David J., Richard C. Elphic, William C. Feldman, Herbert O. Funsten, and Thomas H. Prettyman. (2010). "Performance of Orbital Neutron Instruments for Spatially Resolved Hydrogen Measurements of Airless Planetary Bodies." *Astrobiology* 10 (2): 183–200.

Leinweber, G., D. P. Barry, M. J. Trbovich, J. A. Burke, N. J. Drindak, H. D. Knox, R. V. Ballad, R. C. Block, Y. Danon, and L. I. Severnyak. (2006). "Neutron Capture and Total Cross-Section Measurements and Resonance Parameters of Gadolinium." *Nuclear Science and Engineering* 154 (3). American Nuclear Society: 261–79.

Leitgab, M., Rios, R., Semones, E., Zeitlin, C. (2016). ISS-RAD Fast Neutron Detector (FND) ACO On-Orbit Neutron Dose Equivalent and Energy Spectrum Analysis Status. Paper presented at 21st Workshop on Radiation Monitoring for the International Space Station, 06- 08 September 2016, ESA-ESTEC, Noordwijk, The Netherlands. Retrieved from http://www.wrmiss.org/workshops/twentyfirst/Leitgab.pdf

McGregor, D., Hammig, M., Yang, Y.-H., Gersch, H., and Klann, R. (2003). Design considerations for thin film coated semiconductor thermal neutron detectors—I: basics regarding alpha particle emitting neutron reactive films. Nuclear Instruments and Methods in Physics Research Section A: Accelerators, Spectrometers, Detectors and Associated Equipment, 500(1–3):272 – 308.

McGregor, D. S. and Shultis, J. K. (2011). Reporting detection efficiency for semiconductor neutron detectors: A need for a standard. Nuclear Instruments and Methods in Physics16

**Figure Captions**

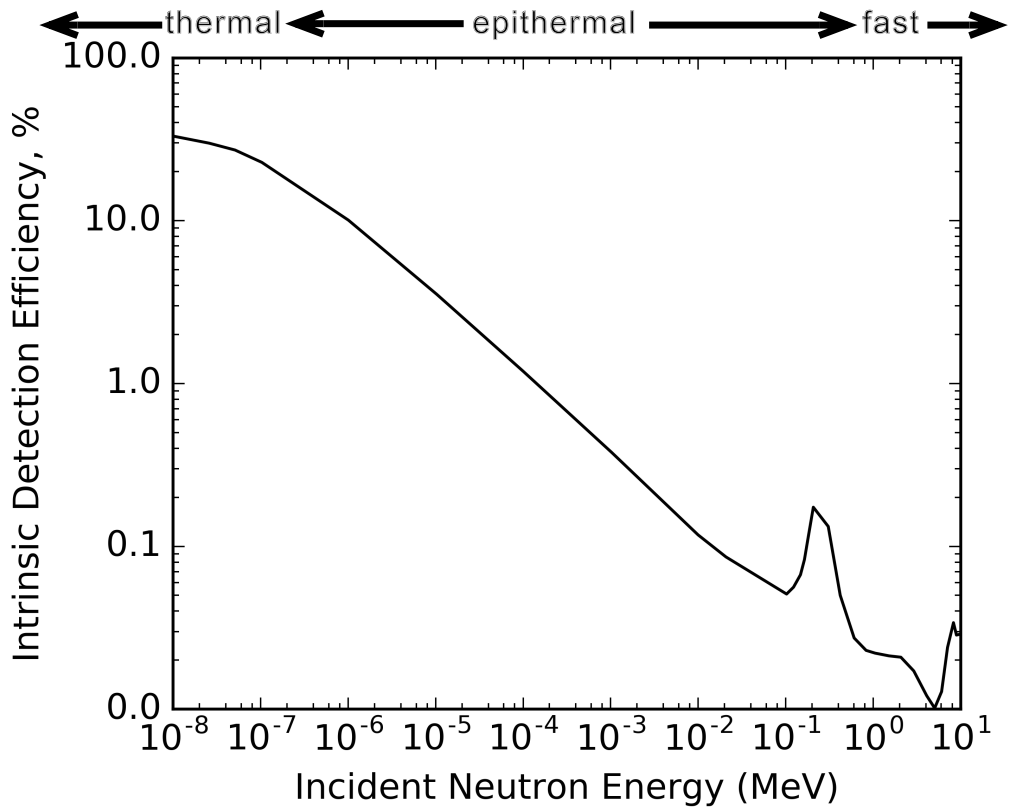

*Figure 1. A simulated MSND energy-dependent neutron response (provided by Radiation Detection Technologies, Inc.).*

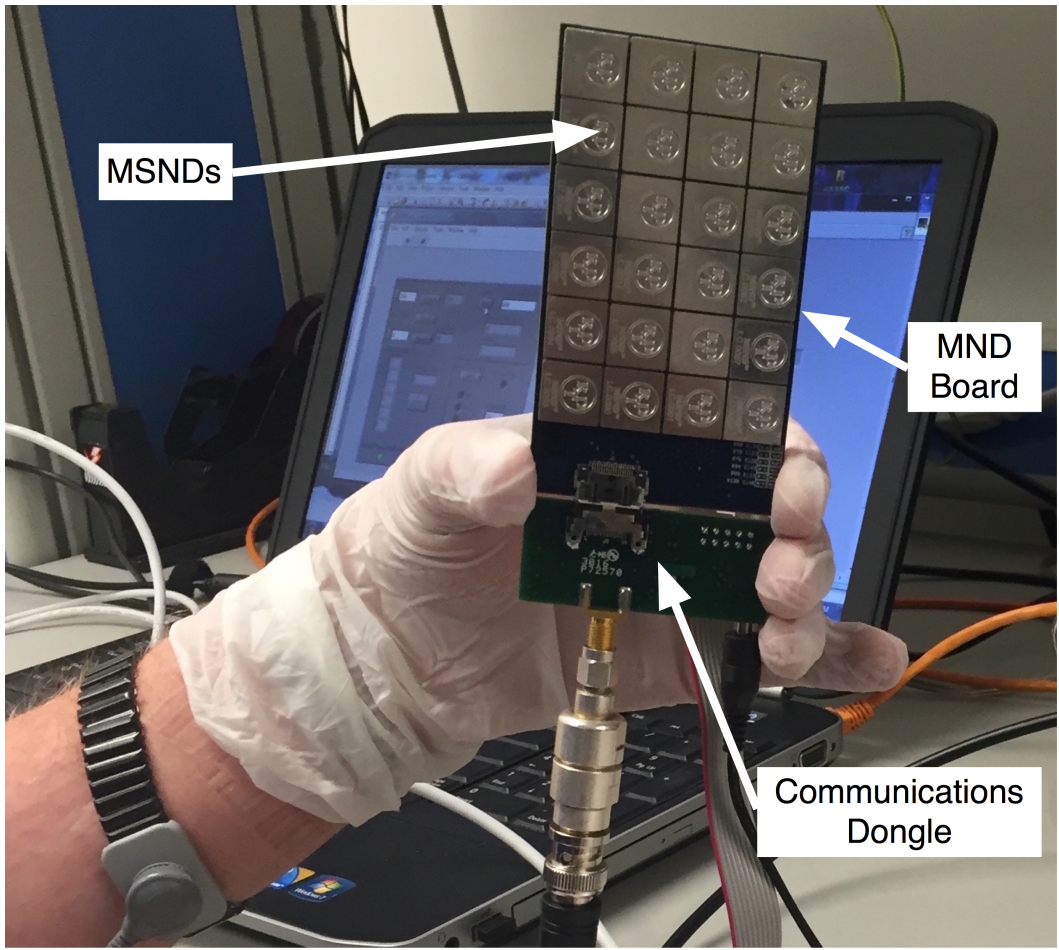

*Figure 2. Shown is a single Modular Neutron Detector (MND) board populated with a 4 x 6 array of MSNDs. The 24 MSNDs are bonded onto an electronics board that provides basic readout electronics. The I2C output signal from the MND is propagated to the counting electronics via a second readout board attached to the bottom. This interface board provides communications with the rest of the instrument electronics.*

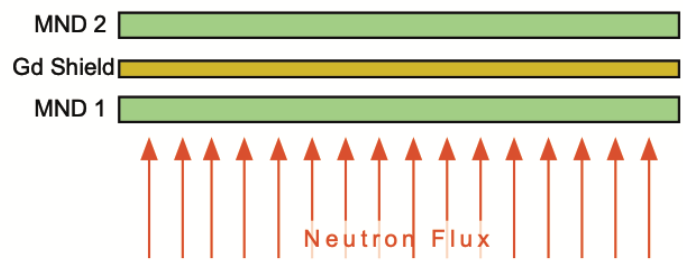

*Figure 3. A schematic of configuration #1, in which a Gd sheet is placed between two MND detectors. The Gd sheet acts as a filter, so that MND 1 measures thermal and epithermal neutrons while MND 2 measures epithermal neutrons.*

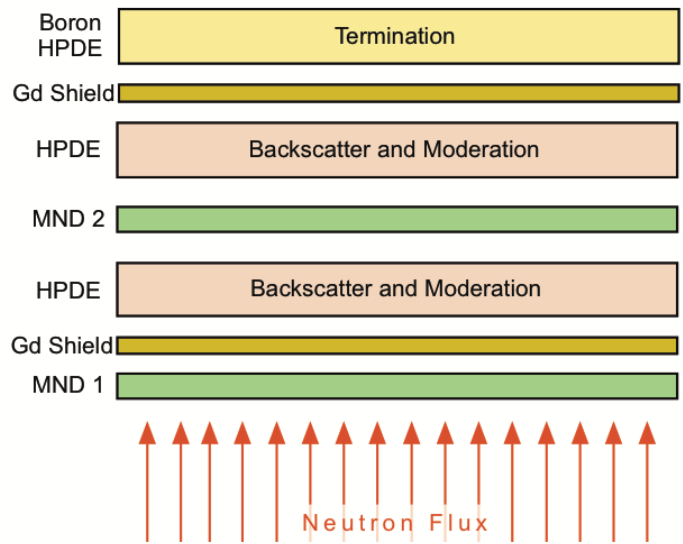

*Figure 4. A schematic of configuration #2. This configuration uses two types of HDPE, undoped HDPE and boron-doped HDPE, to provide backscattering, moderation, and absorption of the neutrons.*

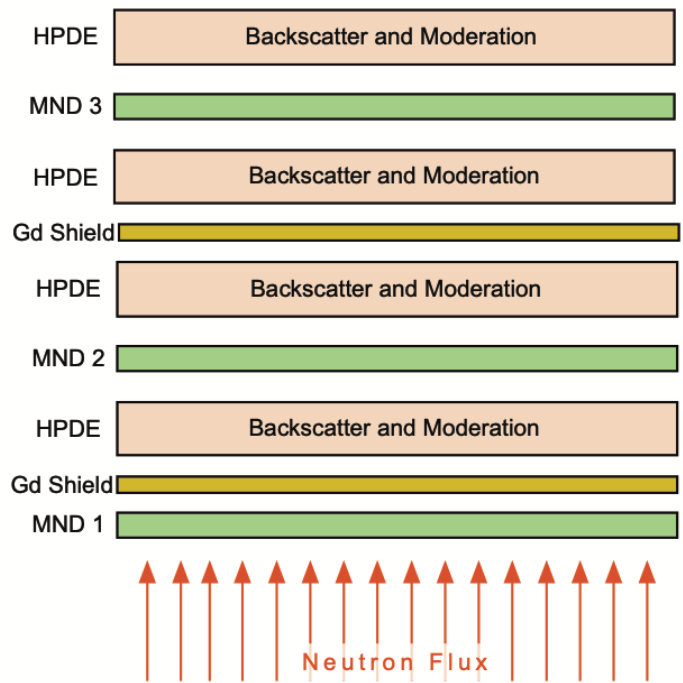

*Figure 5. A schematic of configuration #3. This configuration uses only undoped HDPE to provide backscattering and moderation of the neutrons, particularly the epithermal neutrons. A third detector board is added to increase the number of neutrons counted in this configuration.*

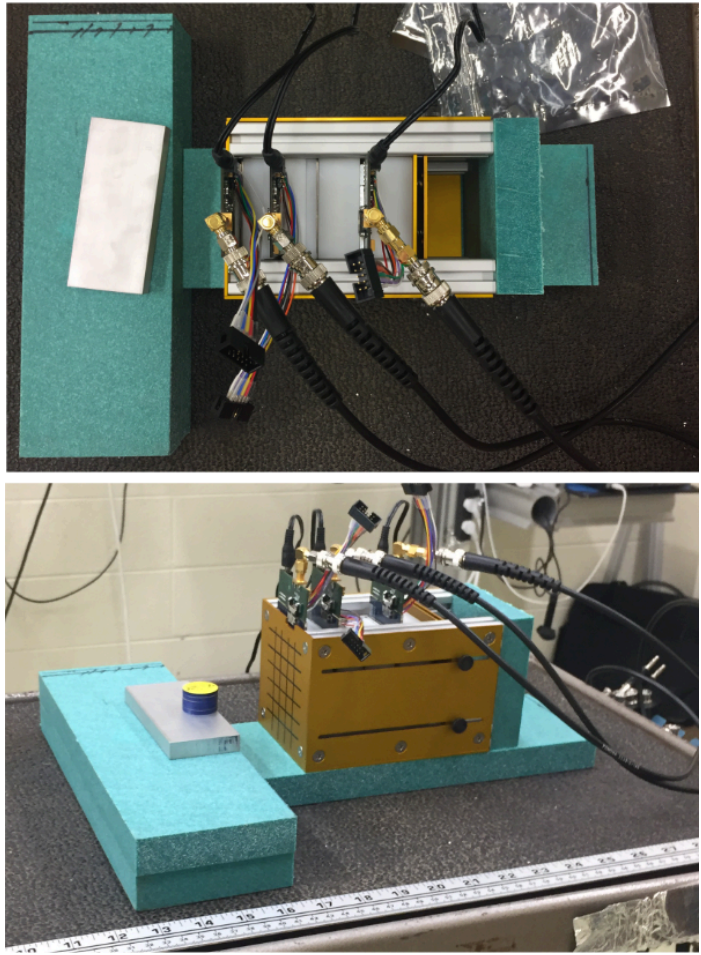

*Figure 6. Shown (top) is a top down view of the prototype instrument in Configuration #3. The aluminum chassis is anodized in gold, while the white blocks within are the HDPE blocks. The blue/green blocks are external pieces of borated HDPE, as discussed in the text. The MND boards are wedged between the HDPE blocks. Shown (bottom) is the setup for one of the radiation tests for Configuration #3. The data acquisition boards attached to the MNDs can be seen jutting out of the prototype chassis.*

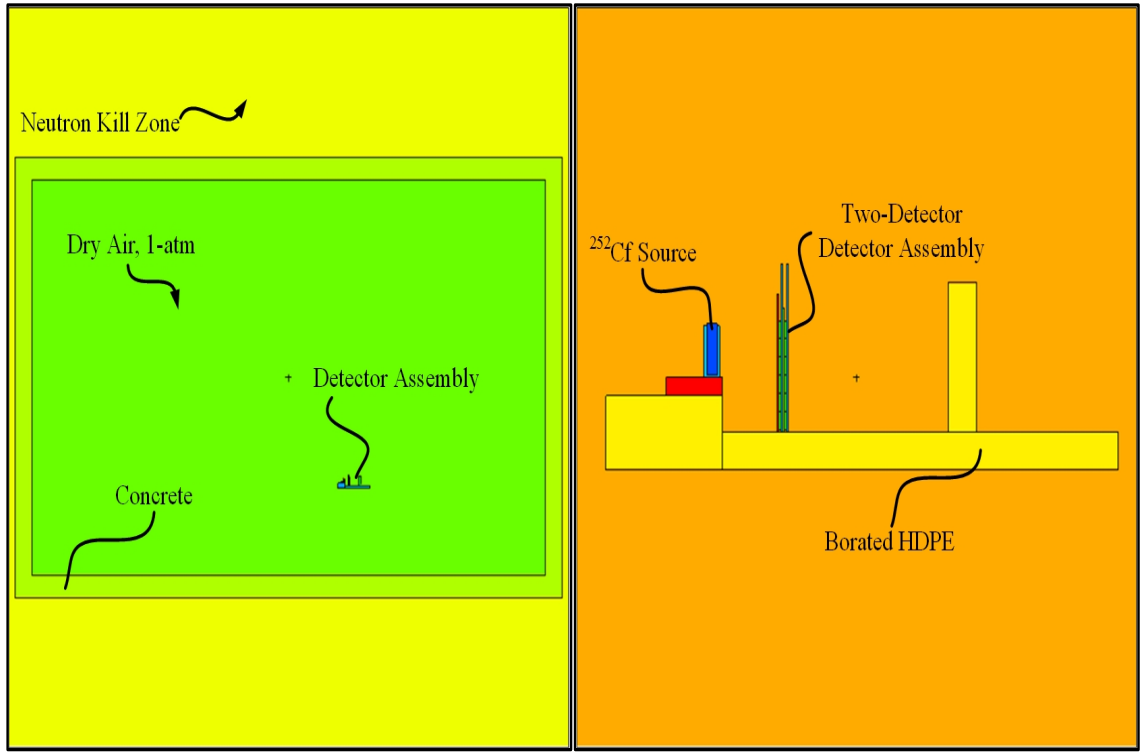

*Figure 7. Shown (left) is a virtual view of the room modeled after the room in which all measurements were conducted. The model did not include external equipment or personnel present during the measurement periods. Shown (right) is a close-up view of one of the detector assemblies, showcasing the arrangement of borated HDPE.*

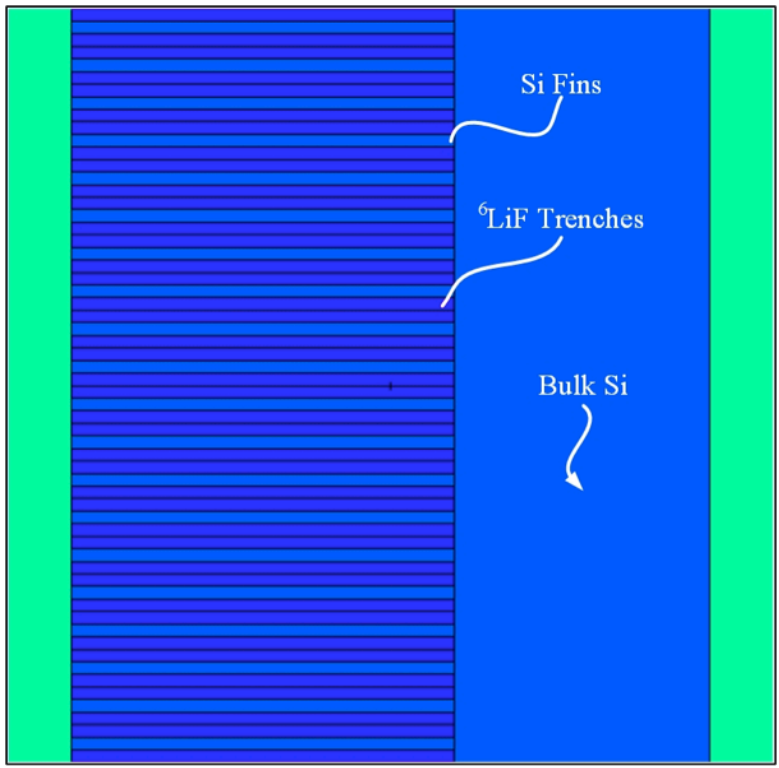

*Figure 8. Depicted is a cross-sectional view of an MSND modeled in MCNP6. Trenches are etched and backfilled into natural Si with $^6$LiF . Neutrons absorbed in the $^6$LiF are converted to the proper charged-particle reaction products which are transported through the device. Energy deposited into the Si is tallied.*

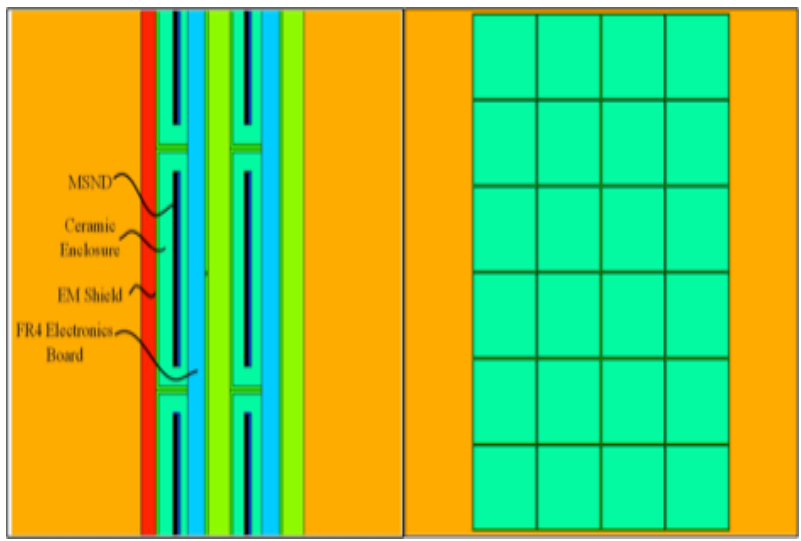

*Figure 9. (left) Depicted is a cross-sectional view of a stack of Modular Neutron Detectors (MNDs). Each MND is modeled as an array of twenty-four MSNDs arranged in a 4 x 6 array (shown right). MSNDs are encased in a ceramic disposable board (CBD) and coated with a layer of electromagnetic (EM) shielding.*

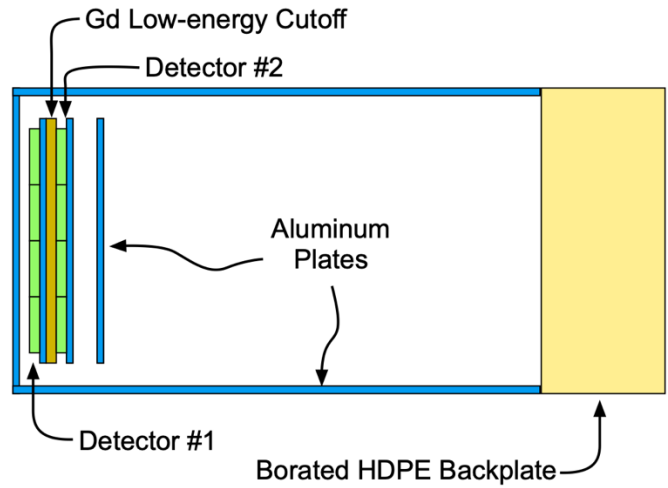

*Figure 10. Depicted is a top-down view of the 2-Detector assembly (Configuration #1) as modeled in MCNP6.*

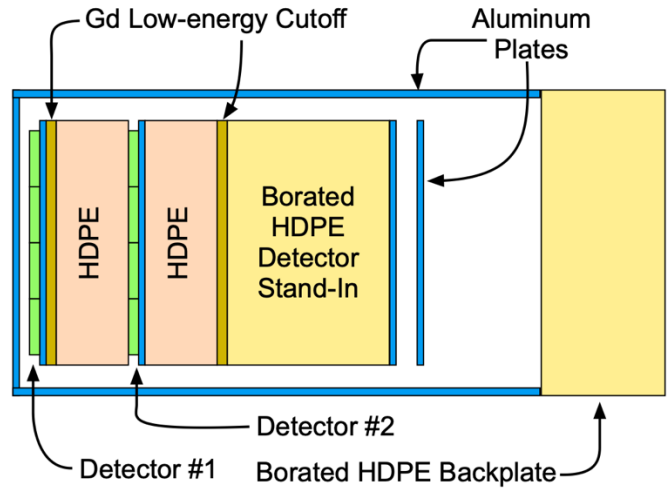

*Figure 11. Depicted is a top-down view of Configuration #2 as modeled in MCNP6.*

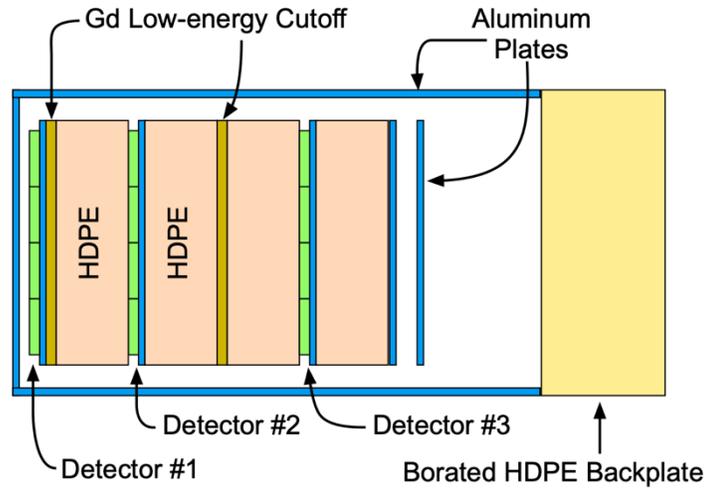

*Figure 12. Depicted is a top-down view Configuration #3 as modeled in MCNP6.*

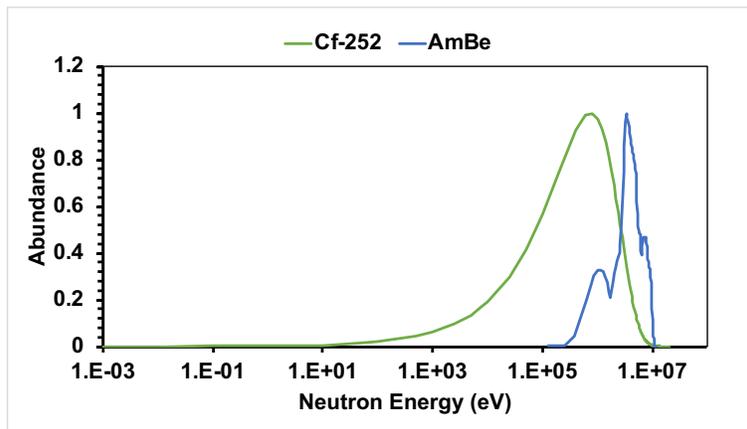

*Figure 13. The simulated relative abundance of neutrons normalized to the most probable energy emission for the two sources, as modeled in MCNP6. The $^{252}$Cf source was modeled using a Watt fission spectrum (Watt, 1952), and the AmBe source energy distribution was calculated using SOURCES4C (Shores, 2002).*

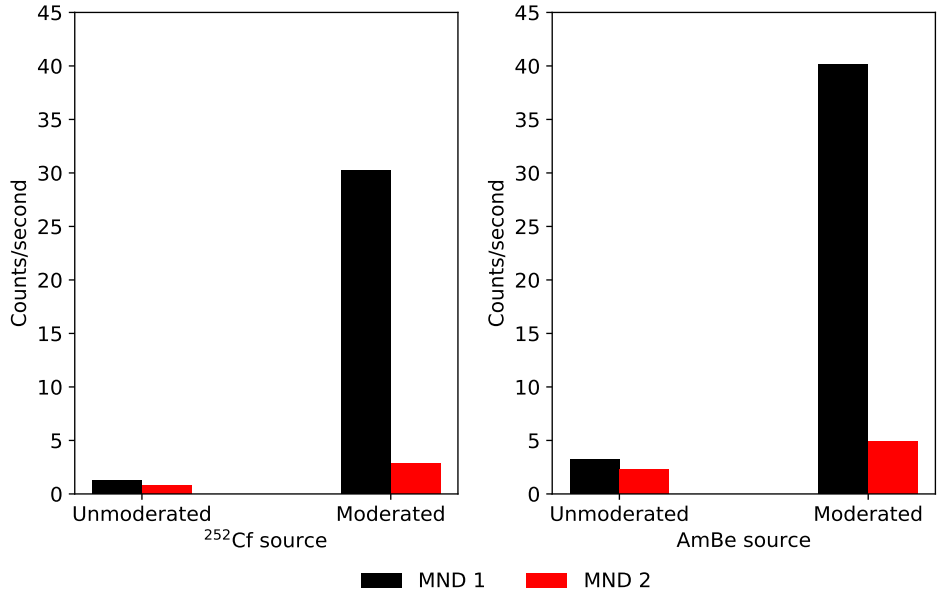

*Figure 14. Neutron response to different sources for configuration #1. The use of a moderated source increases the number of neutrons detected by the first detector (MND 1), improving the statistics for the experiment.*

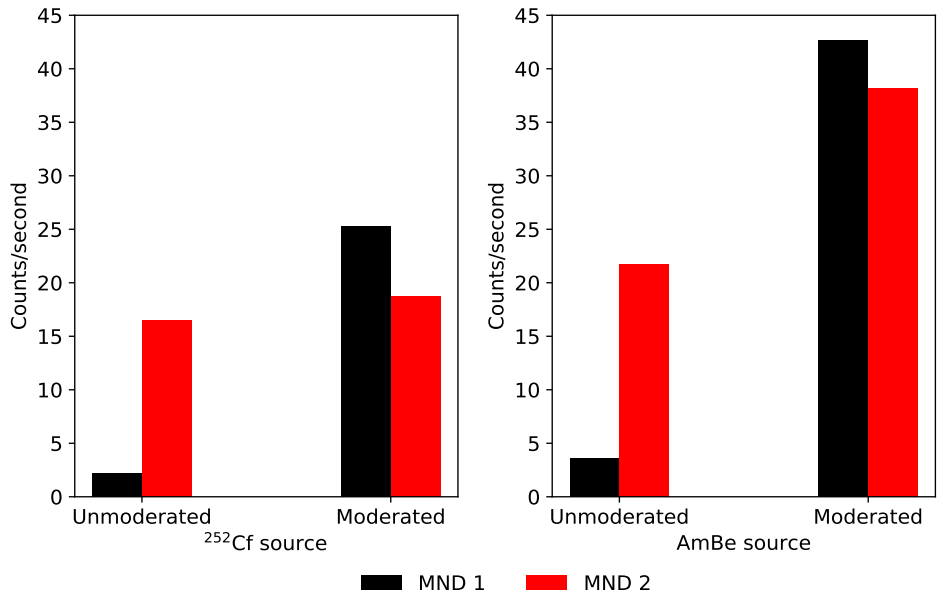

*Figure 15. Neutron response due to different sources for configuration #2. The addition of HDPE after the Gd shield has clearly increased the neutron detection in the MND 2.*

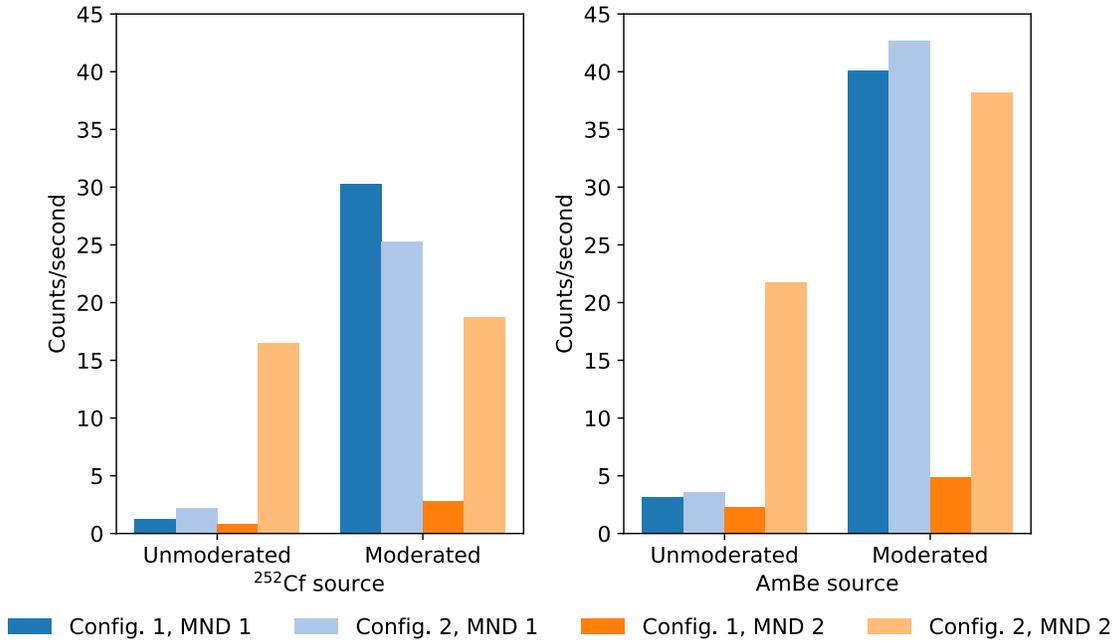

*Figure 16. Comparing the neutron count rates between Configuration #1 and #2 for both the unmoderated and moderated sources.*

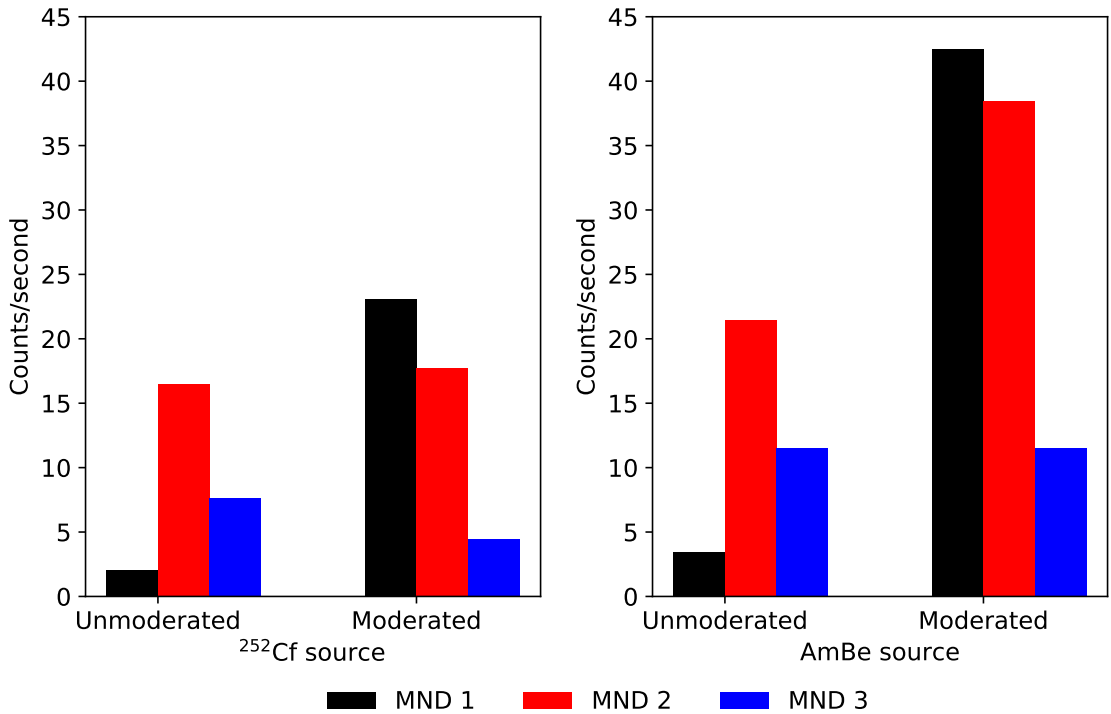

*Figure 17. Neutron detections due to different sources for configuration #3.*

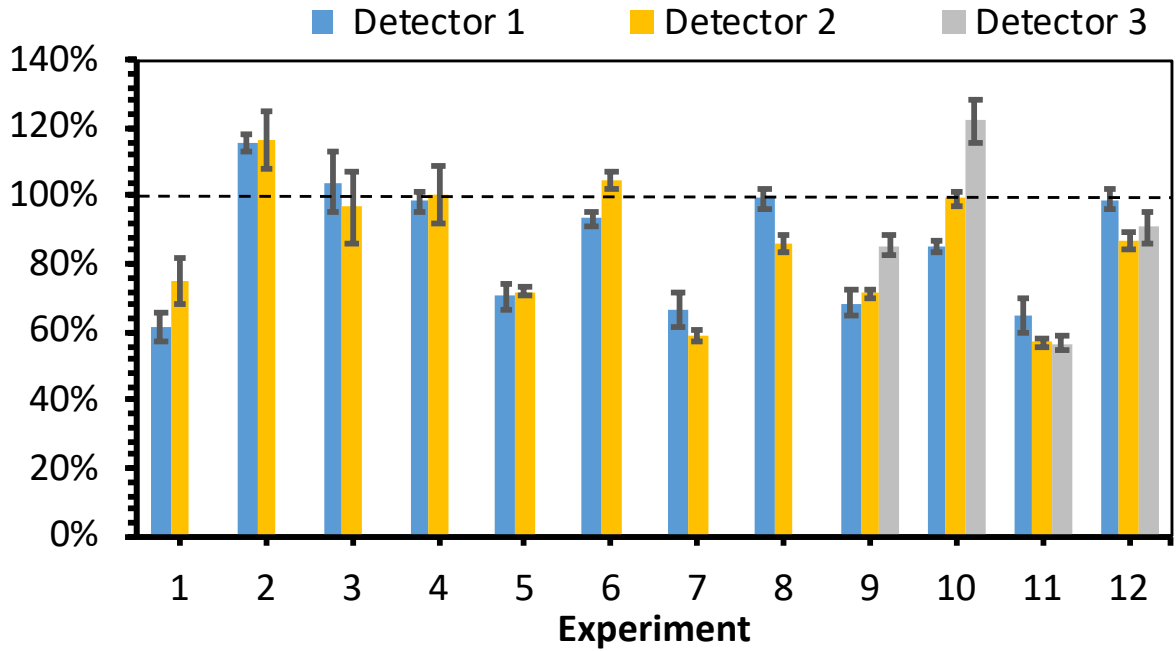

*Figure 18. Plotted are the ratios of the measured count rates to the modeled count rates for each MND channel. The dashed line marks where the measured and modeled counts match. Note that only the last four experiments included a third detector.*

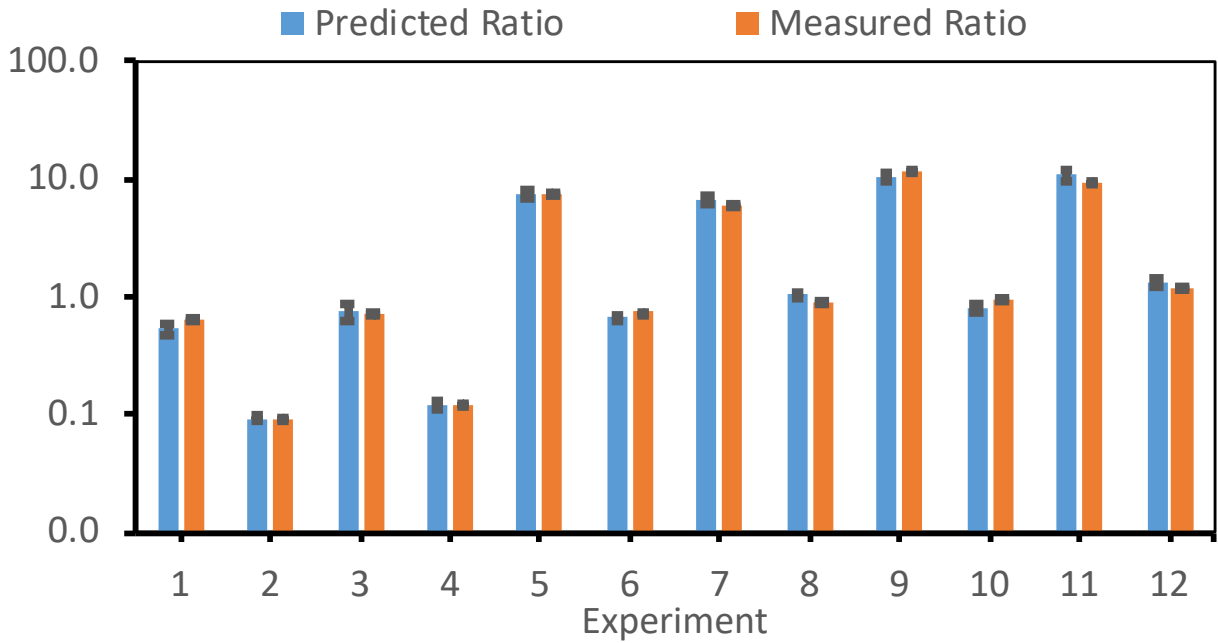

*Figure 19. Plotted are the ratios of the neutron detections by MND 2 and MND 3 to the neutron detections by MND 1 for each scenario for both measured and modeled results. Models were in good agreement with the measured results.*